# Study of the light nuclei cluster structures
# in three-body photodisintegration reactions.
# Application for the excited states of $^6$He nucleus

N.Demekhina, H.Hakobyan, Zh.Manukyan, A.Sirunyan and H.Vartapetian
A.I.Alikhanyan National Science Laboratory
(Yerevan Physics Institute)

## Abstract

An experimental program for the study of the cluster structures of excited states for the light nuclei of He, Li and Be in three-body photodisintegration processes is proposed. The investigations will be realized with $^6$Li, $^7$Li and $^9$Be targets in seven photodisintegration reactions. As an application, the photodisintegration of $^7$Li nuclei into the (t+t+p) final state with excitation and decay of neutron-rich $^6$He nucleus into the (t+t) channel is considered. The full Monte-Carlo simulation of the experiment is done with a special impact on the study of experimental setup performance, concerning the photon and excited states energy resolutions. The experiment will be performed on the bremsstrahlung photon beam of the Yerevan Electron Synchrotron.

## 1. Introduction

Numerous experimental and theoretical works are being carried out today which are devoted to the study of light nuclei [1].

The structure of excited states is of an increasing interest from both, experimental [2-6,13] and theoretical [7-9] points of view, corresponding to the existence and manifestation of the cluster composition of those states. Clustering is a general phenomenon observed not only in stable light nuclei but also in short living nuclei, such as $^6$He and $^{11}$Li [1].

In most of the cases excited states are recognized as nuclear molecular resonances in a two-cluster system (or one cluster and nucleon "N") in which one of the clusters might be also excited. The coexistence of different cluster structures is also possible, leading to different decay channels [7-9]. There is a continuum of unbound states above the break-up threshold and exact determination of their excitation energies, widths and modes of decay is very important in choosing and building the adequate theoretical models.

For the experimental study of excited states, the proton and ion beams [3-5,13], as well as the photon [2] and $\pi$-meson beams [6] are mostly used. There is much inconsistency in the present experimental data on the excitation spectra of the lightest nuclei. This is in particular valid for the excited states of $^6$He (identification of energy levels and decay modes) made by a few experimental groups.

The most recent data on the $^6$He energy levels obtained with the ion beam of $^7$Li [3,5] disagree with those measured with π-meson beam [6] on the number of excited states of $^6$He as well as in their energies and widths, that might be explained [6] by high energy resolution and strong suppression of the physical background. But probably this cannot be the main cause, as the experimental results [5,10] concerning the excited states in the light nuclei $^6$He, $^6$Li and $^6$Be are in accord and show the expected presence, in the above mentioned isotopes, of excited states of "three – nucleon cluster" with energy: $E_x=18,0\pm1$MeV $\Gamma=9.5\pm1$MeV for $^6$He$^*$(t,t); $E_x=18.0\pm0.5$MeV $\Gamma=5\pm0.5$MeV for $^6$Li$^*$(t,$^3$He) and $E_x=18,0\pm1.2$MeV $\Gamma=9.2\pm1.3$ MeV for $^6$Be$^*$($^3$He,$^3$He).

The data obtained with ion beams in the ion-ion collision processes $^6$Li ($^7$Li, $^7$Be) $^6$He, $^6$Li($^3$H,$^3$He) $^6$He (see references in [2]) and that obtained with a photon beam $^7$Li (γ,p)$^6$He [2] also disagree, which can be explained not only by complicated background contributions in ion-ion processes, but also by the many possible combinations of angular momentum transfer between the projectile and the target [2].

In this work we present a program for investigation of the cluster structure of excited states of light nuclei. It concerns the photodisintegration processes of $^6$Li, $^7$Li and $^9$Be target-nucleus in three –body final states [11]. By selecting the final states, consisting of clusters and nucleons the physical background in these reactions is to some extent known, which simplifies the analyses (see below).

The program gives the possibility of investigating three types of excited states cluster structures (mostly not studied yet). The general $A^*$(C1,C2) excited states, where C1=cluster, C2=cluster (the case of $^6$He$^*$(t+t)), or C2=N. There are also two other types of cluster structures; the first one A(C1$^*$,C2)$^*$ (or A(C$^*$,N)$^*$), for example, the excited state of the parent nuclei A(C,N) consisting of excited cluster C$^*$ and nucleon N might be the case of $^7$Li nuclei, containing an excited $^6$He$^*$ and proton, and the second one is the case with excited states of the type A(C1,C2,N)$^*$, where C1=cluster, C2=cluster or nucleon (also the case of $^7$Li (t+t+p)$^*$).

As an application, the complete Monte-Carlo simulation and the analysis of the three - body final state reaction γ+$^7$Li → t+t+p is presented, aimed at investigating the excitation energy levels of $^6$He in the process of γ+$^7$Li → $^6$He$^*$+p, followed by the decay of $^6$He$^*$→ t+t.

In analogy with γ+$^7$Li → t+t+p reaction and the application for the excited states of $^6$He nucleus, it is possible to investigate the cluster structure ($^6$He$^*$+p)$^*$ and (t+t+p)$^*$ of the excited states in $^7$Li nucleus and also the three types of the excited states in six isotopes of the light nuclei, namely $^5$He, $^5$Li, $^6$Li, $^7$Li, $^8$Be, $^9$Be in three-body final state reactions with $^6$Li, $^7$Li and $^9$Be targets.

## 2. The program of the excited structures study

In this study (see also [11]) we consider the photo-disintegration reactions with three-body final states γ+A → 1+2+3, where the particles (1, 2, 3) in general are the nucleons (p, n) and light nuclei (d, t, $^3$He, $^4$He(α)). When the targets are $^6$Li, $^7$Li and $^9$Be, the particles (1, 2) are (p, d, t, $^3$He, α), while the (3) one is the nucleon (p, n) [11].

In these conditions we observe the following seven photo-disintegration reactions:

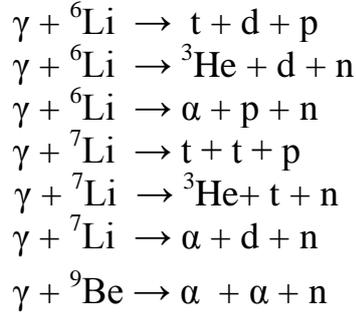

$$\gamma + {}^6Li \rightarrow t + d + p$$
$$\gamma + {}^6Li \rightarrow {}^3He + d + n$$
$$\gamma + {}^6Li \rightarrow \alpha + p + n$$
$$\gamma + {}^7Li \rightarrow t + t + p$$
$$\gamma + {}^7Li \rightarrow {}^3He + t + n$$
$$\gamma + {}^7Li \rightarrow \alpha + d + n$$
$$\gamma + {}^9Be \rightarrow \alpha + \alpha + n$$

There are three types of decay channels for these 7 reactions of photodisintegration:

- A statistical channel $\gamma + A \rightarrow 1+2+N$ which represents a physical background channel of the reaction ($\gamma + A$) and which can be calculated as a three-body phase-space.
- The channels of production and decay of the excited isotope states in the given reaction $\gamma + A \rightarrow 1+2+N$ and precisely in our case in the form of three two-body processes: $\gamma + A \rightarrow (12)+N$, $\gamma + A \rightarrow (1N)+2$ and $\gamma + A \rightarrow (2N)+1$ with 3 resonances' excited states, $B^* = (12)^*$, $(1N)^*$, $(2N)^*$. They are decaying in the final three-particle states.
- The channels of production and decay of the 3 particles excited states $(1+2+N)^*$.

For the study of these seven reactions $\gamma + A \rightarrow 1+2+N$, the production angles and the kinetic energy of the known particles (1, 2) will be measured in coincidence in two detectors. These measurements are able to determine: the energy ($E_\gamma$) of incident photon on the target, the parameters of excited state (x) decay products (particles 2 and 3), invariant mass or excitation energy ($E_x$) and width ($\Gamma_x$) of the excited state [11].

For these seven photo-disintegration reactions we present the cluster structures of 22 excited states of seven isotopes: $^5$He, $^6$He, $^5$Li, $^6$Li, $^7$Li, $^8$Be, $^9$Be as well as the targets being used:

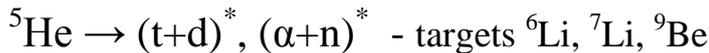

$^5$He $\rightarrow$ (t+d)$^*$, (α+n)$^*$ - targets $^6$Li, $^7$Li, $^9$Be

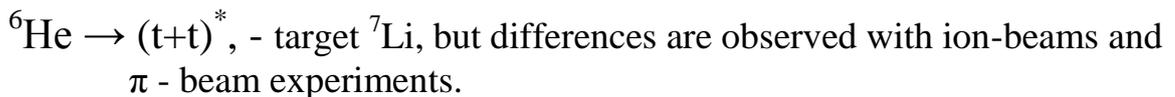

$^6$He $\rightarrow$ (t+t)$^*$, - target $^7$Li, but differences are observed with ion-beams and
    π - beam experiments.

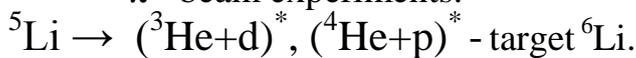

$^5$Li $\rightarrow$ ($^3$He+d)$^*$, ($^4$He+p)$^*$ - target $^6$Li.

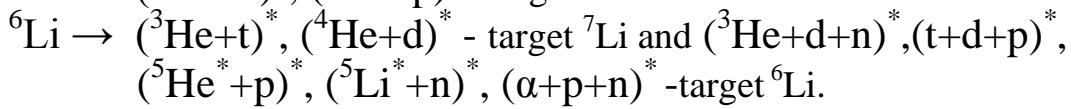

$^6$Li $\rightarrow$ ($^3$He+t)$^*$, ($^4$He+d)$^*$ - target $^7$Li and ($^3$He+d+n)$^*$, (t+d+p)$^*$,
    ($^5$He$^*$+p)$^*$, ($^5$Li$^*$+n)$^*$, (α+p+n)$^*$ -target $^6$Li.

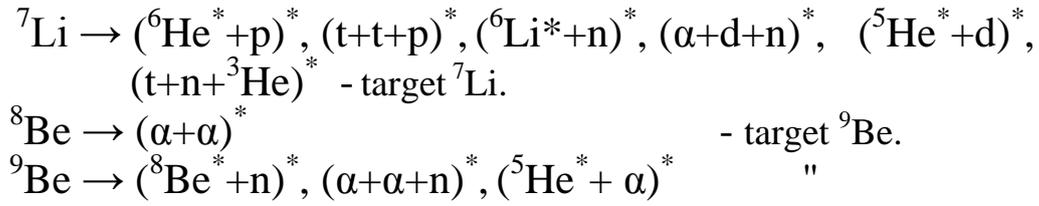

$^7Li \rightarrow (^6He^*+p)^*, (t+t+p)^*, (^6Li^*+n)^*, (\alpha+d+n)^*, (^5He^*+d)^*,$
$(t+n+^3He)^*$ - target $^7Li$.
$^8Be \rightarrow (\alpha+\alpha)^*$ - target $^9Be$.
$^9Be \rightarrow (^8Be^*+n)^*, (\alpha+\alpha+n)^*, (^5He^*+\alpha)^*$ "

The experimental results for a number of excited states presented here have been obtained with different methods [12].

In these seven reactions of type $\gamma +A \rightarrow$ 1+2+N on $^6Li$, $^7Li$ and $^9Be$ targets, where for 6 reactions the particles (1,2) are (d, t, $^3He$, $\alpha$) and only (1,2) are (p, $\alpha$) for the 7th reaction $\gamma + ^6Li \rightarrow \alpha + p + n$, the study of the cluster structure excited state will suffer from the presence of additional background contribution apart of the phase-space one. The data on resonances in (12)$^*$ state will contain the contribution of resonance excitation in [(1N)$^*$+2] and [(2N)$^*$+1] channels, that must be identified and separated from the studied process. It is known, that for these 7 reactions on the 3 targets $^6Li$, $^7Li$, $^9Be$ in 6 of them except the presence of the studied resonance, there is also one additional resonance state (only for the reaction $\gamma + ^7Li \rightarrow ^3He+ t + n$ there are 3 resonance excited states: $^6Li^*(^3He+t)$, $^4He^*(^3He+n)$ and $^4H^*(t+n)$). For determination of the contribution of the additional background resonance ($^4He^*(t+p)$) in the studied reaction $\gamma + ^7Li \rightarrow t + t + p$ (in the Application) see sections 3.2.1 and 3.3.

## 3. Application for the two-cluster (t+t) excited states of $^6He^*$ nucleus

As it was mentioned above, the experimental studies of two-cluster (t+t) excited states of $^6He^*$ nucleus obtained most recently by two various methods ([3,5] and [6]) have given considerably different results (see the data below) for the number, energies and widths of the excited states in the excitation energy range $E_x$=10-40MeV.

**Table 1. Excitation energy levels of $^6He$ discovered in reactions with ions [3, 5] and $\pi^-$ -mesons [6]**

| $^7Li+ ^6Li \rightarrow ^7Be+ (^6He^*(t+t))$ | $\pi^- + ^9Be \rightarrow ^6He^*(t+t)+t$ [6] | |
|---|---|---|
| 18,0±0.5MeV, Γ=7.7±1.1MeV [3] | 15.8±0.6MeV, Γ=1.1±0.6MeV | A |
| 18,0±1.0MeV, Γ=9.5±1.0MeV [5] | 20.9±0.3MeV, Γ=3.2±1.5MeV | B |
| | 31.1±1.0MeV, Γ=6.9±2.3MeV | C |

The compilation data [12] on $^6$He energy levels above 15MeV, obtained so far before 2002 are quite close to that of [6], although with visibly higher widths. These data obtained with ion beams are suffered, as mentioned above, from the presence of multi-particle final states leading to the significant background in (tt) invariant mass spectra. This statement seems is confirmed by the new data recently obtained in the reaction $^3$H($^4$He,tt)p, $^3$H($^4$He,pt)t [13] allowing to observe excitation of narrow $^6$He energy levels ($\Gamma \approx$1MeV) at energies 14.0, 16.1 and 18.0MeV with much lower background.

We believe that the experimental study of the excited states of $^6$He in the energy range $E_x$=10-60MeV, using photons as a probe, is the challenge that will allow us to confirm the already observed states or discover new ones and understand the mechanism of their formation. For this purpose we discuss below the feasibility of the photo-disintegration process $\gamma + ^7$Li $\rightarrow (^6$He$^* +$p$) \rightarrow$ t + t + p study, including the structure and performance of the experimental setup, background conditions and expected rate.

### 3.1. Scheme of experimental setup

The scheme of the experimental layout in the horizontal zx-plane is shown in Fig.1a. The concept of a two-arm set-up of Si-detector telescopes [14] with a sensitive area of (100x100) mm$^2$ for registration of two tritons in coincidence has been chosen. Geometrically the telescopes are located at a distance of 20 cm from the target covering a solid angle of 0.25sr each. Each telescope consists of two perpendicular arrays of 50$\mu$ thick silicon dE/dx sensors and a 1.5 mm thick E-detector with an energy resolution of app.1%, that allows to identify the particles, measure their kinetic energy in the range 4-18MeV and reconstruct the triton angles. The photon beam will be generated by the bremsstrahlung of the 75MeV electron beam of Yerevan Electron Synchrotron [15] with the intensity expected to be above $10^9$ equiv.photons/s (in units of the maximal energy photons). Enriched (99%), 150$\mu$– thick metallic lithium foils will be used. The beam spot size on the target is expected to be less than (10x10) mm2. The target holder and detectors will be mounted in a vacuum chamber. The special target alignment (see Fig.1b) is discussed in section 3.2.2.

### 3.2 Monte Carlo simulation

To work with kinematical distributions and determine the necessary requirements to the performance of the experimental set-up the Monte-Carlo simulations were carried out.

The phase-space Monte-Carlo simulation program [16] was modified to include the resonances in binary system and used to generate the kinematical distributions in:
  - three-body disintegration process (1) - $\gamma + ^7$Li $\rightarrow$ t + t + p
  - quasi-two-body disintegration process (2) - $\gamma + ^7$Li $\rightarrow ^6$He$^*$ + p with subsequent decay of excited $^6$He* states ($^6$He$^* \rightarrow$t+t), and
  - quasi-two-body disintegration process (3) - $\gamma + ^7$Li $\rightarrow ^4$He$^*$ + t with subsequent decay of excited $^4$He* states ($^4$He$^* \rightarrow$ t + p)

The generator [16] was completed by the user package, including the beam-line and setup simulation as well as the kinematical reconstruction and data analysis codes.

As it was mentioned above, the phase-space process (1) represents a general background for the resonance excitation channels, but the resonance process (3) may contribute as a background to the process (2) as well, so the detector configuration should provide a maximal separation of the kinematical distributions of the process (2) and (3) (see below).

There is practically no cross-section data on the process (2) for $^6$He excitation energy above 10MeV, the available ones are not reliable and mostly based on the inclusive or semi-inclusive measurements of the proton or triton yield [17] and thus have not been involved in Monte-Carlo simulations. The possible similarity of the cross-sections in the process (2) and $\gamma + {}^7\text{Li} \rightarrow {}^6\text{He(g.s.)} + p$ [2], is exploited in yield evaluation only (see below).

### 3.2.1 Detector configuration setting

The set of kinematical distributions of the process (2) was simulated with invariant mass $M_{12}$ of two tritons compatible with excitation energies of $^6\text{He}^*$, defined as

$E_x = M_{12} - M_{6He}$, where $M_{6He}$ is the mass of $^6$He ground state and $M_{12}$ can be expressed through the measured kinematical parameters of the tritons as follows:

$$M_{12} = \sqrt{4M_t^2 + 2M_t\left[(T1+T2) - \sqrt{2(T1\times T2)}\times\cos(\vartheta_{12})\right] + 2T1\times T2} \qquad (1)$$

where $M_t$, $T1$, $T2$ are the triton mass and kinetic energies and $\vartheta_{12}$ is the angle between the tritons momenta.

Although the detector registration efficiency for the process (2) certainly depends on the excitation energy of interest, the full energy range $E_x = 10\text{-}60\text{MeV}$ can be covered by two setting of telescopes, optimized for energies $E_x = 20.9$ and $31.1$ MeV [6], including also the choice of the target thickness.

The detector configuration has to provide a compromise between the maximal registration efficiency of process (2) and its kinematical separation from the process (3). For this purpose the telescopes have been set at a non-coplanar configuration ($\varphi_1$-$\varphi_2 < 180^\circ$) that suppresses the contribution of process (3) due to rather coplanar angular distribution of the tritons as compared to the process (2) (see the Fig.2a and insert there). The details of numerical evaluations are presented in section 3.3.

The choice of detector configuration proceeds through the analyses of the polar ($\theta_{t1},\theta_{t2}$) and azimuthal ($\varphi_1,\varphi_2$) angles distributions for the process (2) at fixed photon energy $E_\gamma = 50\text{MeV}$ without limitation on the detector aperture. The Fig.2b shows the distributions of polar angles for the process (2). The telescopes have been installed according to the peak positions in these distributions, allowing to reach the maximal registration efficiency.

The detector configuration parameters: the average polar and azimuthal angles and kinetic energies of the tritons and proton are shown in Table 2.

**Table 2. Setup configuration parameters**

| Configuration | $E_x \pm \Gamma/2$ (MeV) | $<\theta_{t1}>$ (deg) | $<\theta_{t2}>$ (deg) | $<\varphi_1 - \varphi_2>$ (deg) | $<T_{t1,t2}>$ (MeV) | $<\theta_p^{cm}>$ (deg) | $<T_p>$ (MeV) |
|---|---|---|---|---|---|---|---|
| 1 | 20.9±1.6 | 83 | 83 | 120 | 5.7 | 82 | 15.7 |
| 2 | 31.1±3.5 | 87 | 87 | 150 | 9.9 | 76 | 8.1 |

(*) Except $\theta_p^{cm}$, the data are given in the laboratory system.

As one can see from Table 2, the telescopes are installed symmetrically in respect to the photon momentum, they have the same polar angle, but are not co-planar due to proton recoil.

### 3.2.2. Kinematics reconstruction and experimental resolutions calculation

The complete three-body kinematics, including photon and excitation energies $E_\gamma$ and $E_x$, was reconstructed using the six measured parameters: the triton angles and momenta. The expected experimental resolutions can be calculated by comparing them with the true kinematic variables generated for the process (2). The contribution of the experimental uncertainties such as the granularity and energy resolution of detectors, beam spot size, multiple scattering and ionisation energy losses in Li target on the photon and excitation energy resolutions has been simulated and investigated separately, and the results are shown in Table 3 for the particular case of the detector setting optimized for $E_x$=20.9±1.6MeV. During the reconstruction, the average ionization energy losses in the target for the energy of each registered triton are calculated and added to the energy response measured by telescopes. This procedure preserves an appearance of the systematic shift in reconstructed photon and excitation energies. However, apart of the average energy losses in the target, the triton ranges are varying around their average that causes a visible worsening in the excitation and photon energy resolutions. This contribution can be in principle reduced by appropriate target alignment relative to a beam direction, providing correlated and relatively weak variation of the tritons' summary range versus photon interaction depth. The kinematical effect of alignment may be understood from the expression for invariant mass $M_{12}$ (section 3.2.1), while its geometry is shown in the simple case of telescopes co-planarity in Fig.1b. Simple rotation in the reaction plane ($\theta_t = 45^o - (90^o - \theta_D)$) is needed in this case to achieve the necessary alignment. In non-coplanar geometry the situation is more complicated and alignment can be done by two successive rotations. The effect of alignment is however most effective for the telescopes polar angles around $90^o$ and reduced to zero at $45^o$. The results shown in Table 3 are obtained for the alignment calculated for the detector configuration "1".

**Table 3. Contributions of experimental uncertainties on the energy resolutions for configuration "1" in Table 2**

|  | Multiple scattering in target | Uncertainties of ionization losses in target | Beam spot size in target (10x10)mm$^2$ | Detector granularity (10x10)mm$^2$ | Detectors energy resolution | All factors together |
|---|---|---|---|---|---|---|
| $\sigma_{Ex}$ (MeV) | 0.08 | 0.19 | 0.21 | 0.23 | 0.08 | 0.38 |
| $\sigma_{E\gamma}$ (MeV) | 0.47 | 0.28 | 1.22 | 1.28 | 0.08 | 1.71 |

As one can see from Table 3, the main contribution to the excitation and photon energy resolution is coming from the beam spot and detector bin size and lesser from the multiple scattering and variation of ionization losses. The contribution of the detector energy resolution is negligible. As a final result, the account of all experimental uncertainties gives quite an acceptable result as compared with the $E_x$=20.9 MeV level width. The excitation energy resolution has been calculated in the energy range of 10-40MeV for both detector configurations, "1" and "2" (see Table 2) and it runs from app. 0.23MeV at the bottom to a 0.7MeV at the top end of this range.

### 3.3 Simulation results and rate evaluation

- The expected rate of the process (2) for the $^6$He excitation at energies $E_x$=15.8(A), 20.9(B) and 31.1MeV(C) have been evaluated using the experimental data on the process $^7$Li($\gamma$,p)$^6$He(g.s.) [18], obtained for the photon energy range of 50-120MeV and $\theta_p^{cm}$ =24-144$^o$, assuming the same level of the yields for the ground and excited states of $^6$He, certainly indicated by the experimental data [2]. The polar angle distribution for the process (2) in CM system is quite wide, in particular $\theta_p^{cm}$ = 82$^o$ ±18$^o$ / 77$^o$ ±25$^o$ for the detector configuration "1" and "2" (Table 2). The differential cross-sections averaged over the angular and photon energy range accepted for the experimental setup have been evaluated to be in the order of 2-3$\mu$b/sr in the excitation energy range 15-31MeV and photon energy <$E_\gamma$> = 50MeV.

The yield of the process (2) can be calculated according to the standard definition

$N_{evt}$ = $N_t * N_\gamma *$ <d$\sigma$/d$\Omega$> $*$ $\Delta\Omega * \varepsilon$, where <d$\sigma$/d$\Omega$> and $\Delta\Omega$ are the cross-section and simulated proton solid angle in the CM system, while $N_t$, $N_\gamma$, and $\varepsilon$ denote the nuclei density in the target, photon intensity in the acceptance of the setup, and efficiency of registration, that accounts for the detector geometry and particle losses in the target and detector. The efficiency is defined by Monte-Carlo simulation as the ratio of the registered events to the generated ones for the selected range of excitation energy and proton CM angles. The efficiency strongly varies, depending on the excitation energy range, and detector configuration, increasing from 5·10$^{-5}$ at $E_x$=15.8MeV to 0.7·10$^{-3}$ at 20.9MeV and 6·10$^{-3}$ at 31.1MeV, partly due to hardening of the tritons' energy spectra. In particular, at excitation energy $E_x$=15.8MeV it is too soft and partly below the threshold of 3.5-4MeV to reach and be registered in successive layers of Si-pixels (see tritons spectra in Fig.3a and 3b). For the realistic values of $N_t$ and $N_\gamma$ to be in the order of $10^{21}cm^{-2}$ and $10^8 s^{-1}$, respectively, one can expect a maximal event rate of the order of 0.2 event/h, 7.2event/h and 32event/h for the mentioned energies, respectively.

- The invariant mass distribution of two tritons for the processes (2) with excitation energies of $^6$He above 15.8MeV [6] (see Table 1) are plotted in Fig.4a and Fig.4b for detector configurations "1" and "2" (Table 2). Superimposed in the figures are the contributions of the phase-space process (1) as well as not negligible contribution of the process (3) in Fig.4b. The yields of the process (2) and (3) in the figures are consistent with the rate evaluations. As one can see from figures, the yield of the process (2) at excitation energy 20.9MeV is relatively stable in both detector configurations "1" and "2", which are on the average more sensitive to the lower ($E_x$=10-40MeV) and higher ($E_x$=20-60MeV) energy parts of excitation spectra, respectively.

Fig.3a,b and Fig.5a,b show the spectra of the tritons and photons in two detector configurations at the given three excitation energies (Table 1, [6]), allowing to see and evaluate the shapes of the experimental distributions for their use in the data analysis. The contribution of $^6$He excitation at 15.8 MeV is absent in the detector configuration "2", which is explained by its strong kinematical suppression.

The analysis of $\gamma$ -spectra in the reactions $\gamma + {}^7$Li $\rightarrow$ $^6$He$^*$ +p ($^6$He$^* \rightarrow$t+t) and $\gamma + {}^7$Li $\rightarrow$t+t+p may allow us to study the following two types of excited states of $^7$Li$^*$ nucleus like ($^6$He$^*$ + p)$^*$ and (t+t+p)$^*$ (see "Introduction") and determine their energy $E_x$ (connected with certain values $E_\gamma$ of the incident $\gamma$ spectrum).

- The yield of the phase-space process (1) in Fig.4a,b is shown in arbitrary units. According to the observations (see [19] and references therein) and theoretical investigations [20], the mechanism of system's break-up to a three-body final state is prevailed by a sequential decay through the resonances in the binary systems, if they have a width smaller than the excitation energy of the system. As it is indeed the case, one can expect that the phase space contribution will be suppressed in the spectra of invariant mass distribution. This statement is consistent with the experimental data [6] as well.

- The yield at the excitation energy of $E_x$ = 18±4.5MeV [3 and 5] (see Table 1), for which the observations of experiments with ions and $\pi$-mesons are not consistent, was also evaluated to be app. 7event/h, that is similar to that at excitation energy $E_x$ = 20.9±1.6MeV in detector configuration "1".

- The rate evaluation for the process $\gamma + {}^7$Li $\rightarrow$ $^4$He$^*$ + t, using cross-sections data [21] is performed similarly to the process (2), assuming the same magnitude of the differential cross-sections for the ground and excited states of $^4$He for the photon energy range $E_\gamma$ = 30-75MeV. The excitation of $^4$He resonances above 20MeV [22] of the ground state is simulated. The telescopes non-coplanarity alone provides app. two order of magnitude suppression in the yield of the process (3) in configuration "1" as compared to the process (2) at $E_x$=20.9MeV, while in configuration "2" this suppression is approximately of one order of magnitude only at $E_x$=31.1MeV. The shape of (tt) invariant mass distribution in the process (3) (Fig.4b) is quite smooth and similar to that of the process (1), so it seems that the background contribution in configuration "2" would be acceptable for the statistically reliable process (2) separation. Additional kinematical separation of the process (3) contribution is also possible. For example, Fig. 6 shows the proton energy spectra of the process (3) and process (2) at $E_x$=31.1MeV, allowing to see the possibility of simple cut, giving a factor three decrease of the process (3) contribution into the process (2).

- The analysis of the data shown in Fig.4(a) and Fig.4(b) will be done with assumption of a smooth polynomial fit to the invariant mass $M(t1,t2)$ for phase space distribution and Breit-Wigner shapes for the levels excitations. The description of phase-space background, not affected by resonance structures is the subject of a separate study; in particular, it might be built combinatorial, composed of the triton 1 and triton 2 from different events.

- Another important subject of the experimental method is the separation of the accidental background in t+t events produced by different photons within the time gate of registration. With aim to study and remove these contributions, a time-of-flight analysis of the signals registered in two arms can be done. The distribution of the time-of-flight difference measured and reconstructed between two arms of the telescopes can be written as $\delta t = (t1-t2)^{meas} - (t1-t2)^{reco}$. The reconstruction is based on the knowledge of the measured kinetic energies and time-of-flight distances for the detection bins. The true differences are concentrated around $\delta t = 0$ within the width of 1.45ns, consistent with experimental uncertainties of Table 3, while for the accidentals this distribution must be peaked at any beam bunch time, that allows to determine and separate the true process (2) events (see section 3.2).

## 4. Conclusions

We propose to explore the three body break-up of the light nuclei using $^6$Li, $^7$Li and $^9$Be targets in seven reactions of photo-disintegration with two light fragments in coincidence with the purpose of investigating two-cluster resonances excitation in an exotic light nucleus. The use of photons in the study of the cluster structure of the light nuclei would be useful for understanding the mechanisms of the reactions and the related structure of the light nuclei.

The proposed experimental program explores an advantage of the known final state even without detection of the third particle (nucleon) that decreases the uncertainties of the physical background analysis compared to the case of undetected cluster or cluster's fragments.

As an example, the excitation of $^6$He nucleus is considered with a full Monte-Carlo simulation of the reaction $\gamma + {}^7$Li $\rightarrow {}^6$He$^*$(t+t) + p and experimental setup. The physical background, including three body phase space $\gamma + {}^7$Li $\rightarrow$ t+t+p and $\gamma + {}^7$Li $\rightarrow {}^4$He$^*$(t+p) + t reactions was also simulated and possible contribution analysed. As a result the feasibility of the experimental study of the reaction $\gamma + {}^7$Li $\rightarrow {}^6$He$^*$(t+t) +

p is demonstrated, both on experimental resolutions on excitation and photon energies, allowing to resolve and identify $^6$He energy levels and on expected luminosity. According to the yield evaluations, the 250-300 hours beam time would be enough to observe and analyze the excitation of $^6$He levels in the energy range $E_x$=10-40MeV. The same experimental program with variations might be also applied to the list of the photo-reactions presented in section 2.

**Figures captions:**

**Fig.1**: (a) The layout of the experimental setup;
(b) the alignment of the lithium target.

**Fig.2**: (a) The distribution of tritons azimuthal angle difference ($\varphi_1$ - $\varphi_2$) in the process $^7$Li($\gamma$,$^6$He$^*$)p  and  $^7$Li($\gamma$,$^4$He$^*$)t  in the insert;
(b) the triton (t1,t2) polar angle distribution in the process $^7$Li($\gamma$,$^6$He$^*$)p.

**Fig.3**: (a)The tritons spectra simulated for the process $^7$Li($\gamma$,$^6$He$^*$)p at three **r**esonance energies ("A",”B”,”C”) in the detector configuration "1";
(b) the same as in Fig.3a  for two resonance energies  (”B”,”C”),  in  the detector configuration "2".

**Fig.4**: (a)The simulated invariant mass distribution of two tritons for the process $^7$Li($\gamma$,$^6$He$^*$)p for three  ("A",”B”,”C”) resonance energies in the detector configuration "1", superimposed with contribution of phase-space process (1);
(b) the same as in Fig.4a for configuration "2", where the contribution of  $^6$He resonance at $E_x$=15.8MeV is negligible. Shown by dotted line is the contribution of the process $^7$Li($\gamma$,$^4$He$^*$)t.

**Fig.5**:(a) The photons energy spectra simulated for the process $^7$Li($\gamma$,$^6$He$^*$)p at  three resonance energies ("A",”B”,”C”) in the detector configuration "1";
(b) the same as in Fig,5a for two resonance energies (”B”,”C”) in the detector configuration "2".

**Fig.6** The proton energy spectra (laboratory system) for the processes $^7$Li($\gamma$,$^6$He$^*$)p and $^7$Li($\gamma$,$^4$He$^*$)t at $E_x$=31.1MeV.

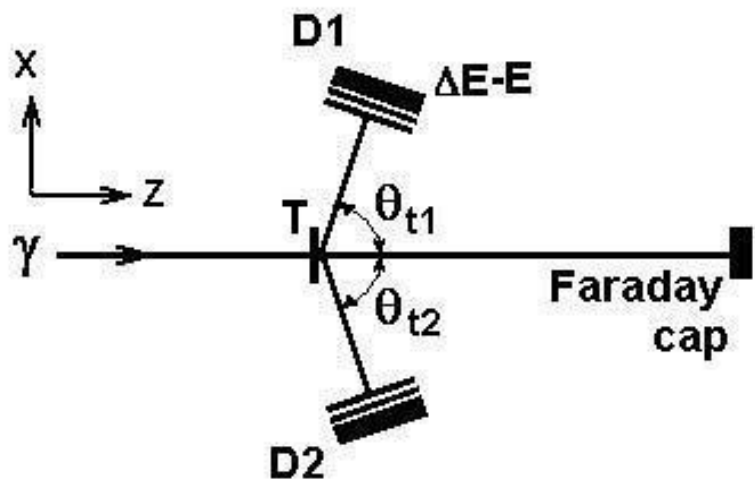

Fig.1a

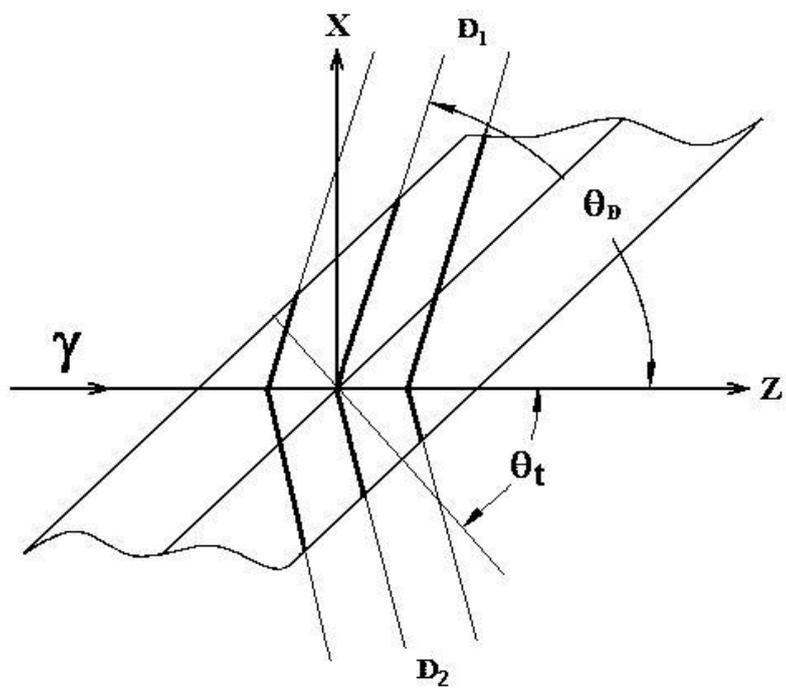

Fig.1b

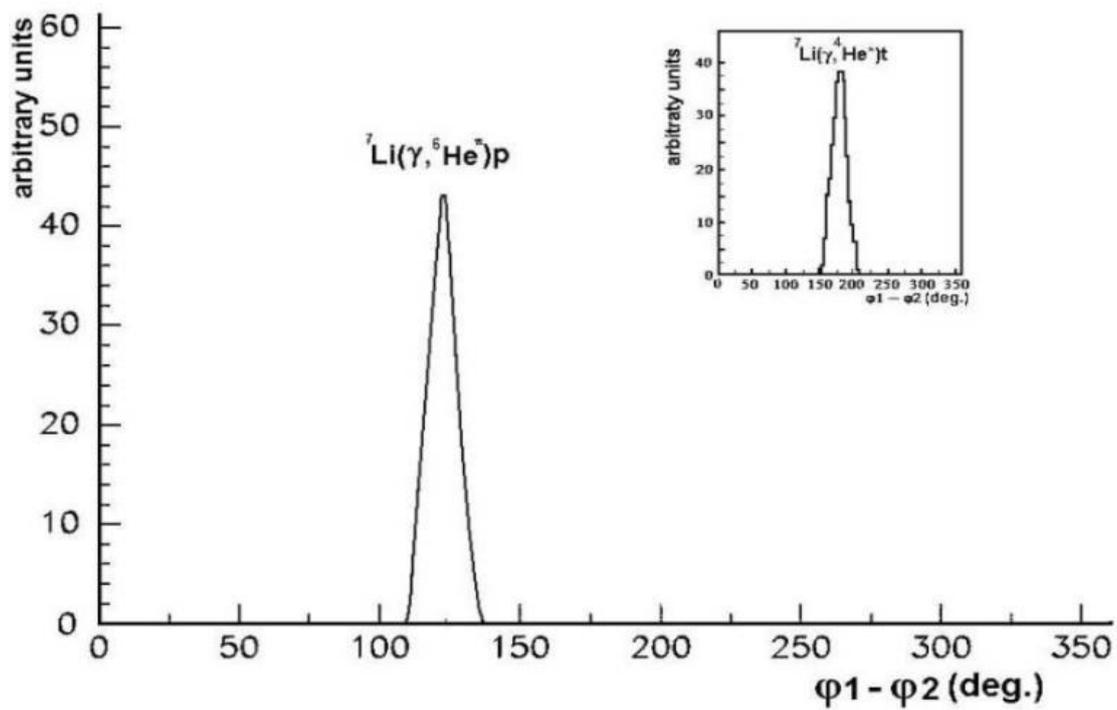

**Fig.2a**

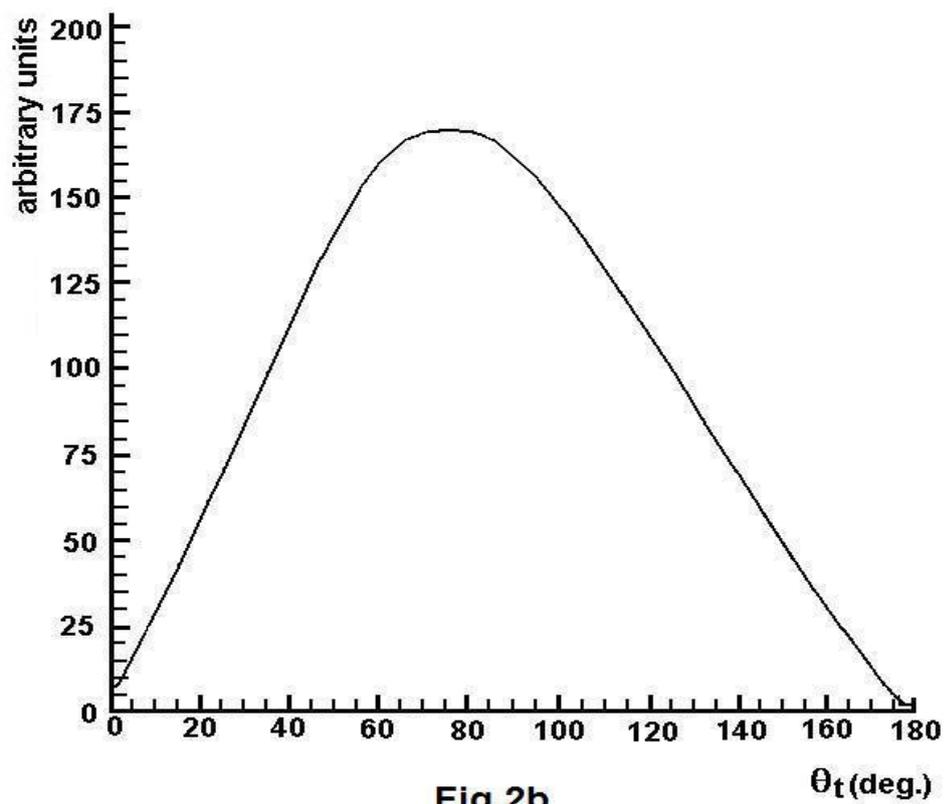

**Fig.2b**

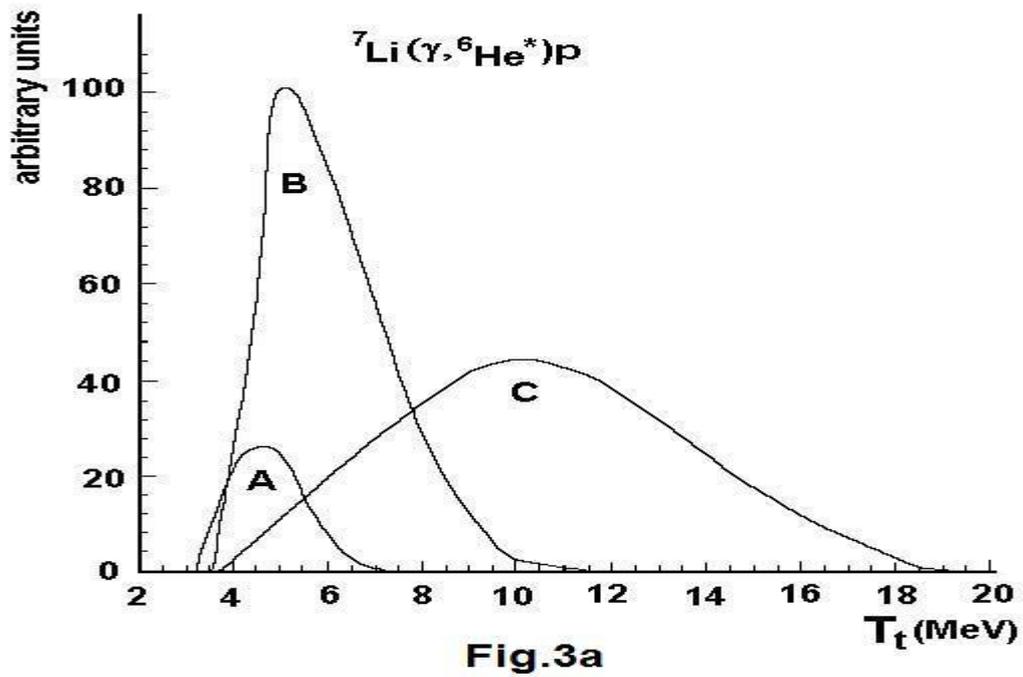

**Fig.3a**

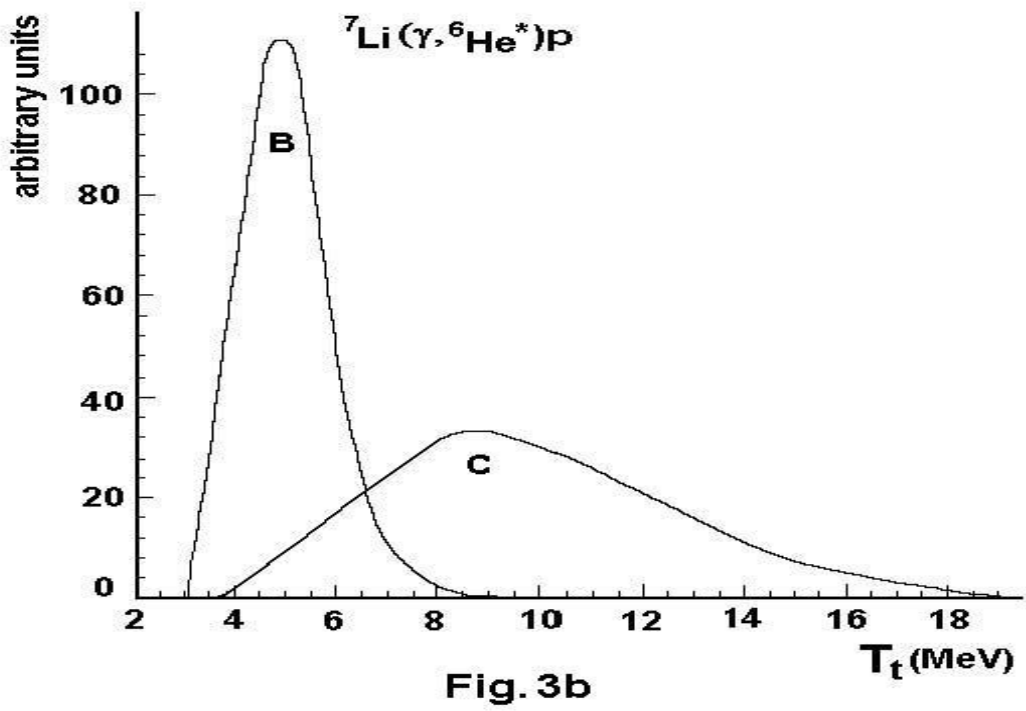

**Fig. 3b**

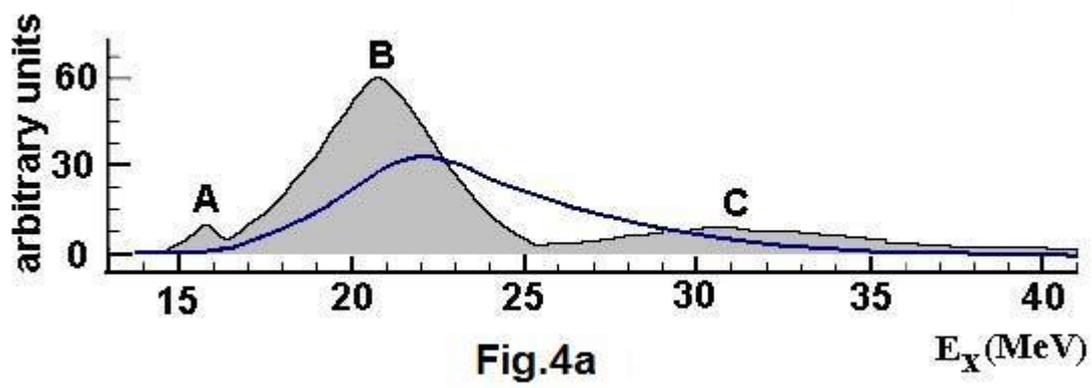

**Fig.4a**

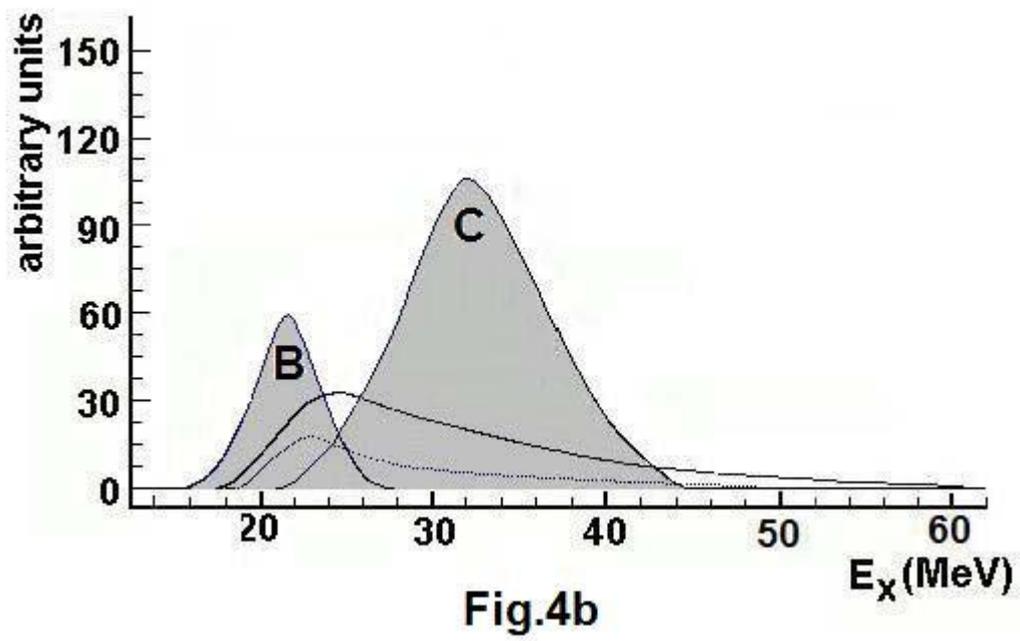

**Fig.4b**

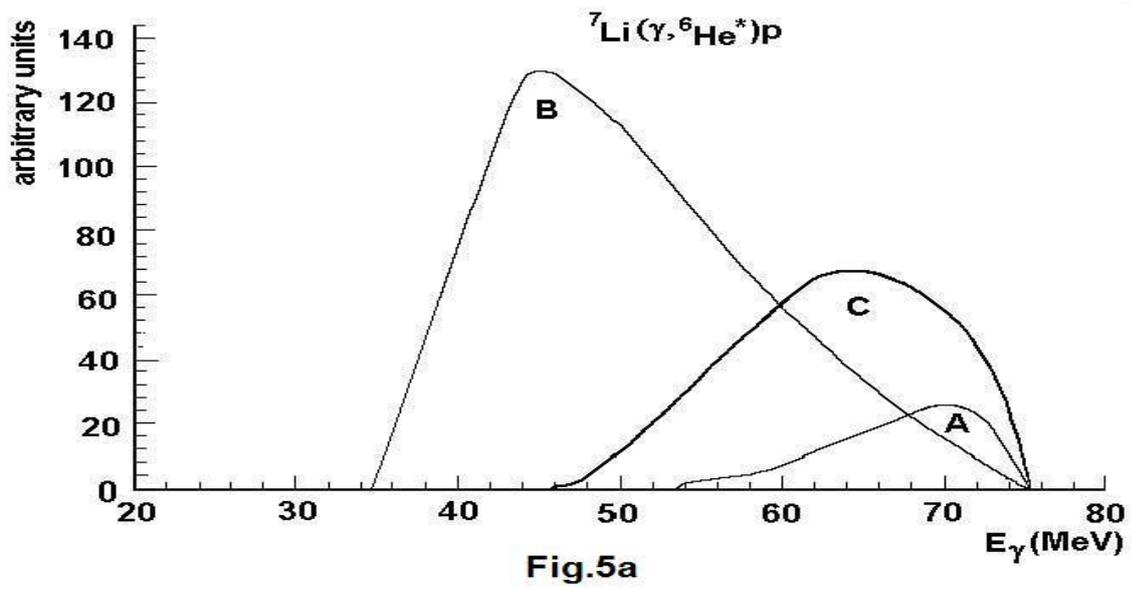

Fig.5a

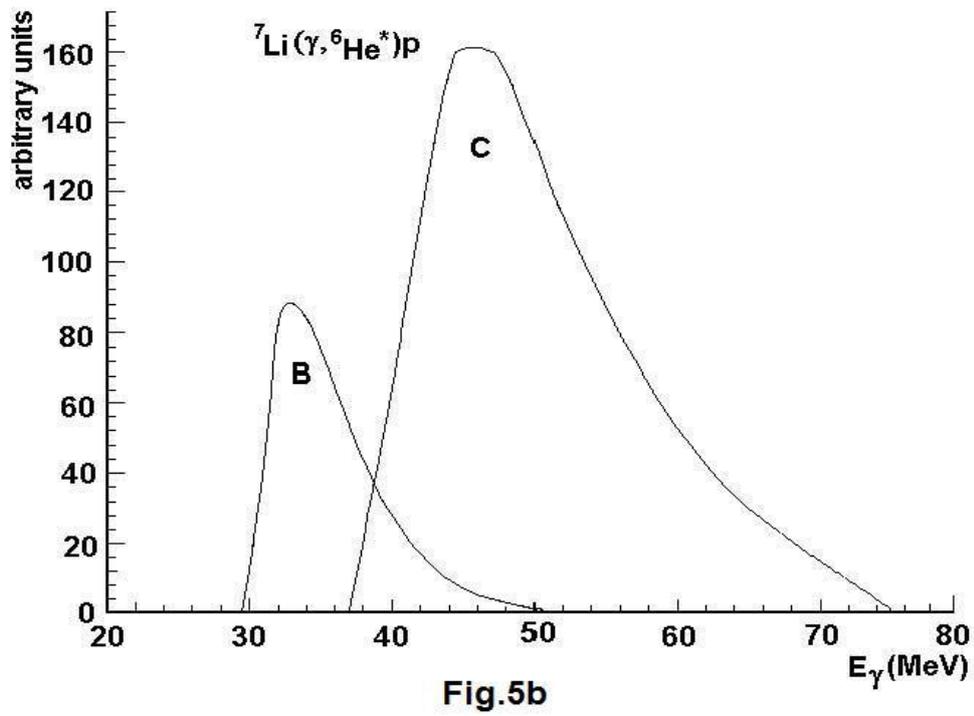

Fig.5b

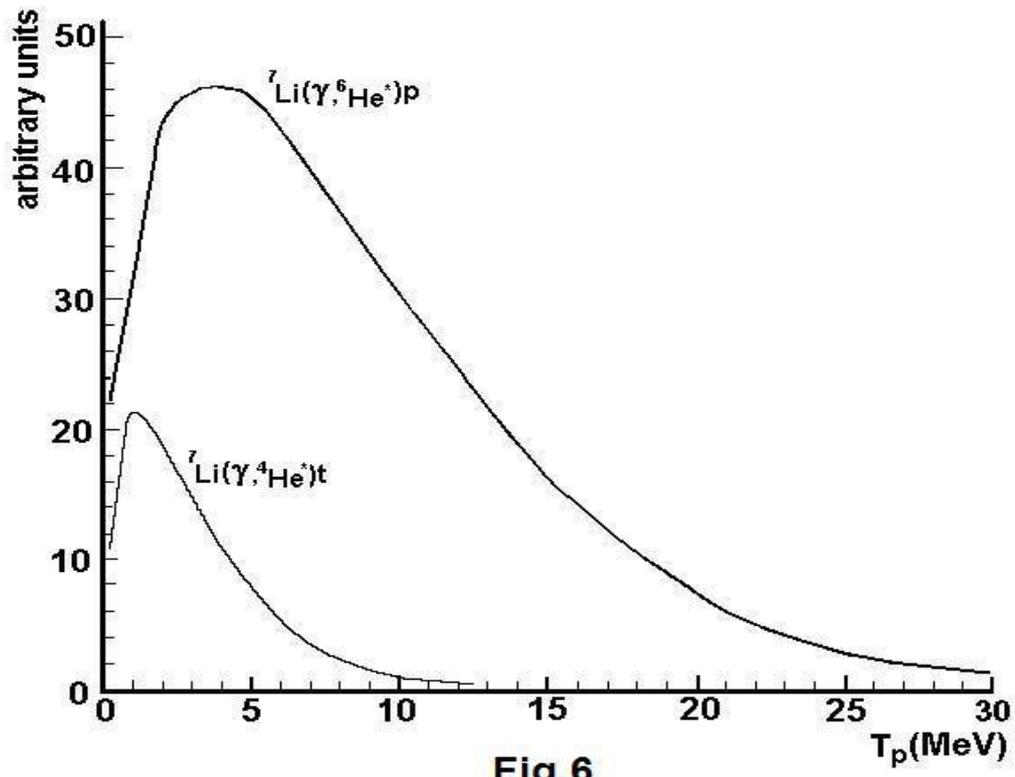

**Fig.6**